\documentclass[10pt]{article}
\usepackage{graphicx}

\begin{document}
\begin{center}
\Large{\bf{SiPM used as fast Photon-Counting Module and for Multiphoton Detection}}
\end{center}
\begin{center}
\bf{P.Eraerds$^1$, M.Legr\'{e}$^1$, A.Rochas$^2$, H.Zbinden$^1$, N.Gisin$^1$}
\end{center}
\begin{center}
\it{$^1$Group of Applied Physics, University of Geneva} \\
Geneva, Switzerland \\
www.gap-optique.unige.ch
\end{center}
\begin{center}
\it{$^2$idQuantique\\ Geneva, Switzerland\\}
www.idquantique.com
\end{center}


\begin{abstract} We demonstrate fast counting and multiphoton detection abilities of a Silicon Photo Multiplier (SiPM). In  fast counting mode we are able to detect two consecutive photons separated by only 2.3 ns corresponding to 430 MHz. The counting efficiency for small optical intensities at $\lambda = 532$ nm was found to be around 8.3\% with a dark count rate of 50 kHz at T$=-7^\circ$ C. Using the SiPM in multiphoton detection mode, we find a good signal discrimination for different numbers of simultaneous detected photons.\end{abstract}

\section{Introduction}
Photon-counting modules are used for more and more applications, for example in high energy physics, quantum communication or fluorescence spectroscopy. Many new techniques have been developed in order to be able to count photons (e.g. avalanche photodiodes (APDs) \cite{Cov}, visible light photon counters \cite{Tak} or superconducting detectors \cite{Golt}). Each technique has its own advantages and drawbacks, so one has to choose the proper technique depending on the requirements of the application. Nowadays, a new demand for high count rate emerges from different fields (e.g. spectroscopy, free space and fiber based quantum cryptography  \cite{The,Takes,Gord,Col,Takes2}, fast Quantum Random Number Generators \cite{Stef, Jenne}, reflectometry \cite{Diam,Legre} or astronomy \cite{Str,Nal}). A solution has been found with NbN superconducting  single photon detectors \cite{Kor}, demonstrating GHz-count rates. However, superconducting photon-counting modules are not as practical as APD based detectors, this is why much effort is made to increase the maximal count rate of APD photon-counting modules \cite{Roc,Giu}.
Presently, photon-counting modules based on silicon avalanche photodiodes (Si-APD) have a count rate limited to about 10-20 MHz (Perkin Elmer, MPD, idQuantique). Notice that classical photomultiplier tubes have maximal count rates of about 3 MHz (Hamamatsu). The limitation of the count rate of APDs is due to a typical deadtime of about 50 ns. It has been proposed to reduce the deadtime by combining several APDs in parallel  \cite{Cas}. The idea is that if we consider for example an array of 100 APDs, then, if one APD has detected a photon, 99 APDs are still active to detect the following photon. It is straight forward to understand that if the photons are distributed over the whole APDs array, the deadtime of the detection module is smaller than the one of a single APD. In  \cite{Cas}, the authors proposed to distribute photons over an array of pigtailed APDs with a very fast switch. This method is very efficient in theory, nevertheless, the loss imposed by currently available optical switches makes it inefficient.\\
In our investigations we pursue the same idea of photon distribution over an array of photodiodes, but instead of using an optical switch we use the  distribution of light at the output of a fiber. As detection device, we use a Silicon Photo Multiplier (SiPM) consisting of 132 independent APDs connected in parallel. In this paper we investigate the fast counting and also the multiphoton detection abilities of a SiPM provided by idQuantique.

\section{Silicon Photo-Multiplier}

\begin{figure}[h]
\centering
\includegraphics[width=6cm]{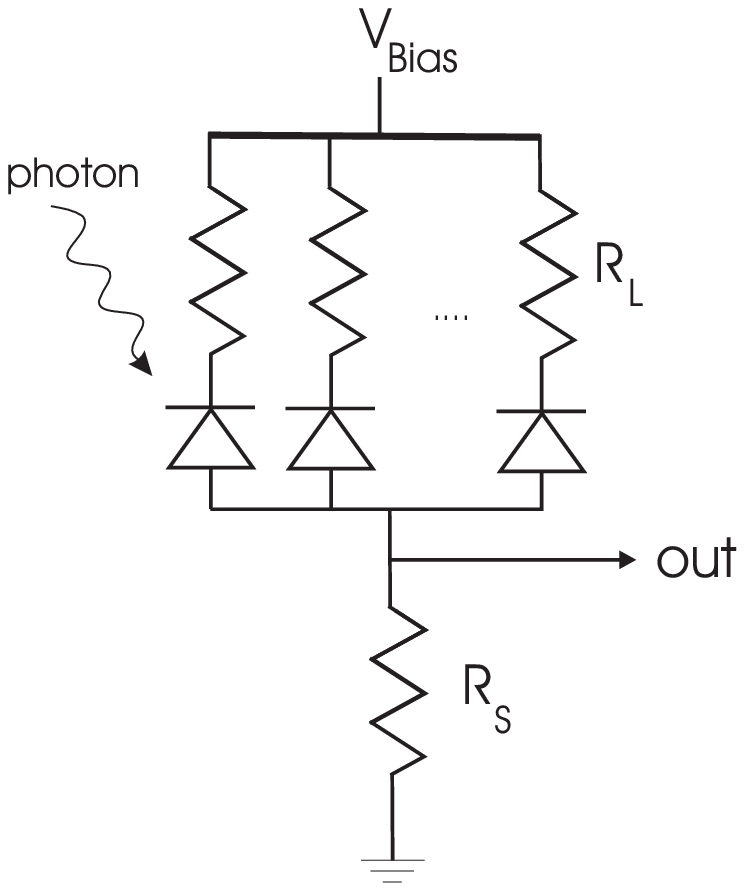} ~~~~~~~
\includegraphics[width=6.5cm]{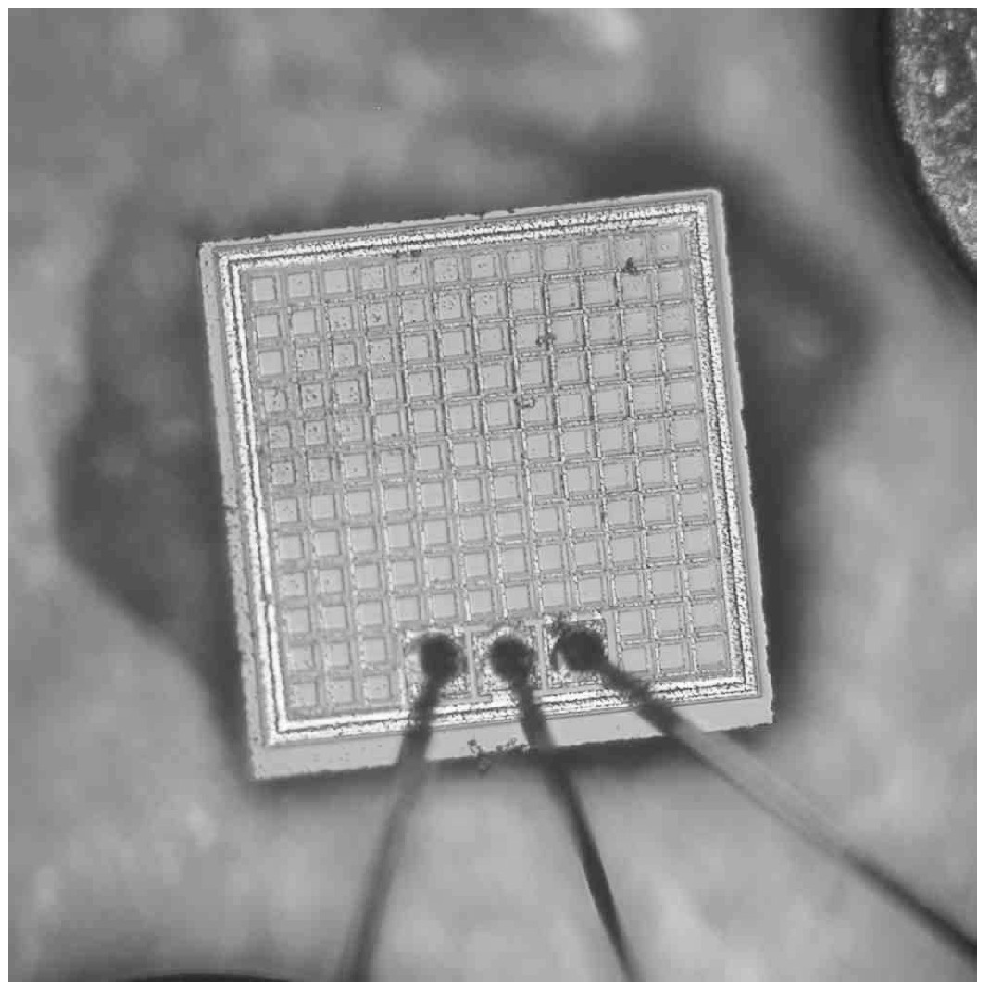}
\caption{Left : Schematic of SiPM. Single APDs are connected in parallel, thus simultaneous detections of single APDs get superposed to form an output signal proportional to the number of simultaneous detections. Right : Photograph of SiPM. 132 pixel (APDs) with a total surface area of 780 $\mu m$ x 780 $\mu m$. A single pixel measures 57.5 $\mu $m x 57.5 $\mu $m.}
\label{SiPMResis}
\end{figure}

For a couple of years, new semiconductor devices called Silicon Photo-multipliers (SiPMs) have been commercially available. They were originally developed to replace classical photomultipliers in nuclear applications \cite{Buz,Gol,Ott} and quite recently they are also used in the bio-medical field \cite{Gri}. Typically, they consist of 100-1000 parallel connected photodiodes mounted on a common chip. A schematic of the electrical design is shown in Fig.\ref{SiPMResis} (left).
We utilize a SiPM consisting of 132 photodiodes like shown in Fig.\ref{SiPMResis} (right). The surface area measures 780 $\mu$m x 780 $\mu$m of which 31 \% (SiPM fill factor) is photosensitive. The breakdown voltage for a single diode is about $V_{bd}=28 $ V and we operate the device in limited Geiger mode around $10\%$ above $V_{bd}$. If there is an avalanche in one of the photodiodes, there is an electrical pulse at the output of the SiPM. But, if two or more avalanches occur at the same time, the amplitude of the electrical output pulse is equal to the number of avalanches times the amplitude of a single avalanche. So, the height of the output pulse is proportional to the number of detected photons - in first approximation, if we neglect the noise.

\section{Multiphoton Detection}
\begin{figure}[t]
\centering
\includegraphics[width=6.5cm]{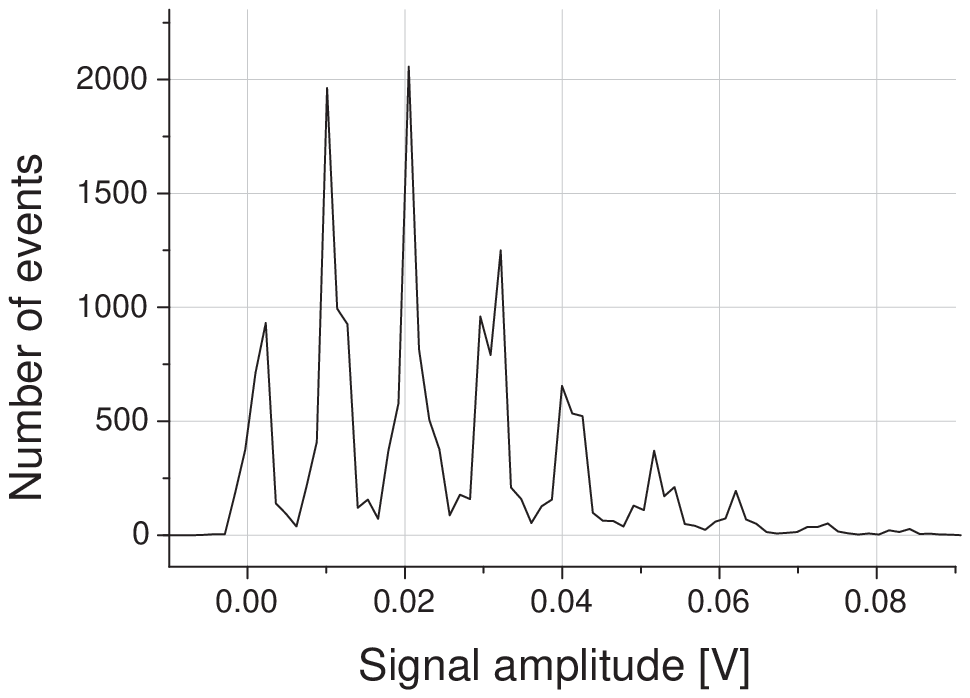}  
\includegraphics[width=8cm]{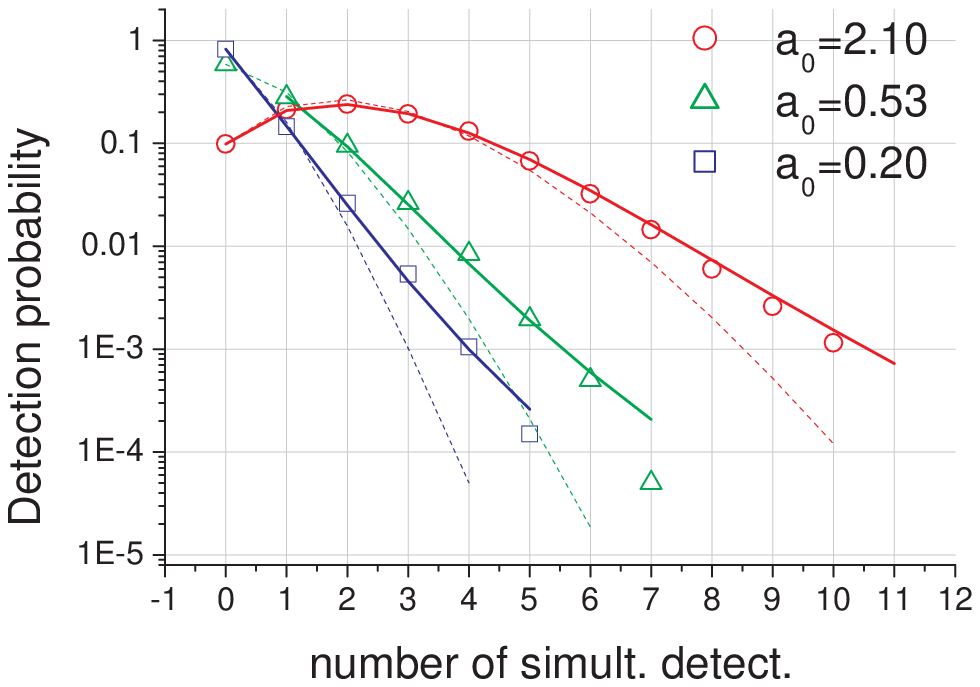}
\caption{Left : Distribution of output signal heights of SiPM when illuminated with weak coherent optical pulses with a fixed average number of photons per pulse. The first peak corresponds to no detection, the second peak to a number of simultaneous detections equal to 1 and so on. Right : Probability of obtaining a certain number of simultaneous detections (see left) for three different mean numbers of incident photons/pulse (hollow points). Values $a_0$ represent the mean number of these data distributions. Dashed curves are Poisson distributions having the same mean number as the data distributions. Solid curves are obtained with Eq.(1) accounting for cross-talk effects.}
\label{MultiPh}
\end{figure}

We start by characterizing our device in multiphoton detection configuration, which is the way it is commonly used. Since the electrical output pulse is very small, we use an electrical amplifier (Mini-Circuits, 0.1-500 MHz) to increase the signal. When we send weak coherent optical pulses on the SiPM, we observe electrical pulses with heights variing from one pulse to another. Their distribution is shown in Fig.\ref{MultiPh} (left). The amplitudes of the output pulses are measured with an oscilloscope which is triggered with the driver of the optical source. Hence, we can measure detections as well as the absence of  a detection. The first peak in Fig.\ref{MultiPh} (left) corresponds to no detection (amplitude $\approx 0$ mV). The second peak corresponds to one detection, the third to two simultaneous detections and so on.
Here we see that it is possible to distinguish the different heights corresponding to the different numbers of simultaneous detections. Hence our device is able to resolve multiphoton events. From this, we  determine the probability distribution shown in Fig.\ref{MultiPh} (right), here for three different mean numbers of  simultaneous detections $a_0$. Because we are sending coherent pulses, we expect to obtain  Poissonian distributions. For each distribution we extract the mean number of detections from the probability of getting no detection. We then plot the Poissonian distributions associated to these mean numbers, shown on the same graph (dashed curves). As we can see, they underestimate the measured values when probabilities of detection numbers are very low. This deviation can be explained by an effect of cross-talk between two adjacent APDs. This means, that a detection in an APD creates an avalanche in one of its neighbours. Considering a very simple model (to first order), we compute the following redistribution of the probabilities:
\begin{equation}
p(n)=\frac{p_{th}(n)+(n-1)p(n-1)p_{ct}}{1+np_{ct}}~,~~ n>0 
\end{equation}~
$$
p(0)=p_{th}(0)$$
where: $p(n)$ is the probability to get n simultaneous detections, $p_{th}(n)$ : theoretical probability (Poissonian\footnote{In this case, using coherent pulses.}) of getting n detections (in the absence of cross-talk) and $p_{ct}$ : probability to get an avalanche due to a cross-talk effect.  The probability distributions obtained with this model, using a cross-talk probability of 9.7 \%, are represented by the solid curves in Fig.\ref{MultiPh} (right). We determined the cross-talk probability by taking the first two values of the distribution with $a_0=0.2$ . This approximation agrees well with the measured data points.


\section{Setup for Fast Counting}

\begin{figure}[t]
\centering
\includegraphics[width=6.5cm]{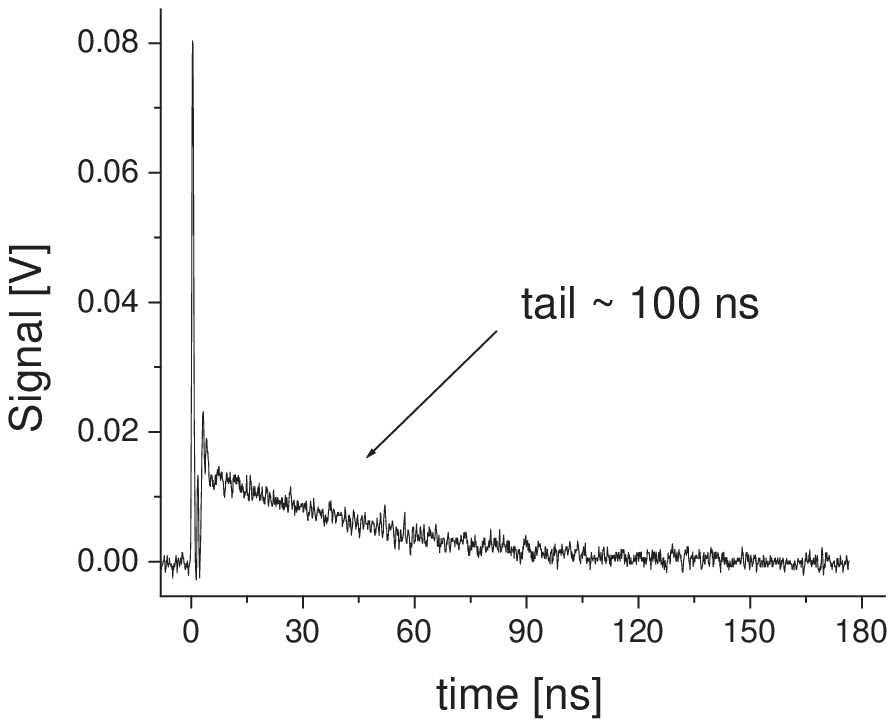}
\includegraphics[width=7.5cm]{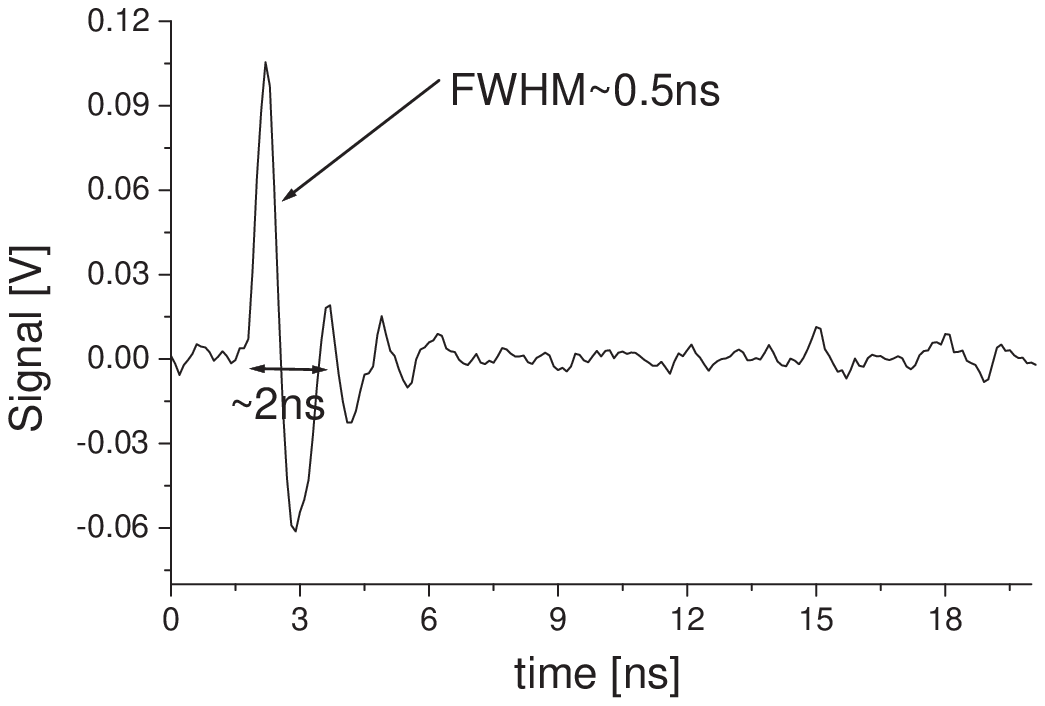}
\caption{Left : SiPM output signal in multiphoton counting configuration. Right : Improved output signal behind high pass filter, suited for fast counting. (Signals represent the minimal response due to a single incident photon)} 
\label{singlesignal}
\end{figure}

To use the SiPM as a fast counting detector the output signal has to be sufficiently narrow to be able to distinguish between consecutive counting events on a small time scale. The output signal Fig.\ref{singlesignal} (left), which we use to detect multiphoton events, is inadequate since the tail of the signal would rapidly add up at high signal rates. Then, the level of the actual detection peak gets higher with each detection and it is impossible to use a threshold criteria.
We decide to use a high frequency pass filter to make the derivative of the signal. The output signal obtained is shown in Fig.\ref{singlesignal} (right). As can be seen, only the narrow peak of the previous signal is kept. The detection peak has a height of ~110 mV with a full width half maximum of about 0.5 ns. The electrical noise is around 15 mV. Because of the negative ripple which happens around 1 ns after the peak, we can expect to distinguish two detection peaks separated by more than 2 ns corresponding to a maximal detection rate of about 500 MHz.\\
\begin{figure}[b]
\centering
\includegraphics[width=12cm]{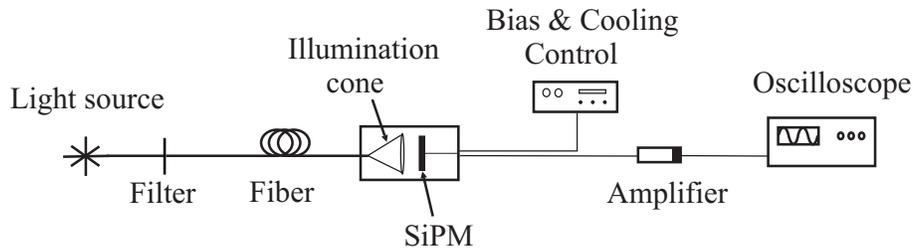}
\caption{Setup for fast counting. The light signal is sent onto the SiPM via fiber. The illumination is intended to be homogenously distributed over the SiPM using the natural distribution of light at the output of the fiber. The output signal is amplified and analysed with an oscilloscope. The number of detections is determined by counting the number of detection peaks in the oscillocope data.}
\label{setup}
\end{figure}
The setup optimized for fast counting is shown in Fig.\ref{setup}. We use different light sources, suited for different regimes of signal repetition rates.  To obtain a certain variety of input powers we use several combinations of grey filters.\\
Behind these filters the light signal is coupled into a fiber and gets distributed over the SiPM at its end. The SiPM itself is mounted on a Peltier element and is housed in a small plastic box to shield ambient light. After passing the amplifier (Miteq, 1-2 GHz) the output signal is detected with an 6 GHz oscilloscope (LeCroy Wavemaster 8600A). The count rate is then determined by a simple program searching for peaks above a certain threshold (40 mV). With this kind of unusual setup we try to mimic standard discrimination electronics. For the moment we do not have an electrical discriminator able to count such relatively weak and short pulses.

\section{Measurement}
In the first place we study general characteristics, i.e. noise and efficiency depending on bias voltage and temperature. Then a first test of fast counting abilities compares our SiPM and a standard single APD with optical pulse repetition rates from 1 to 40 MHz. In a second step we investigate the behaviour when repetition rates are close to the expected limit of 500 MHz. For that purpose, we use a mode-locked laser with a pulse repetition rate of 430 MHz.
The final measurement takes a look at the response to a cw source.

\subsection{Noise and Efficiency}
\begin{figure}[h]
\centering
\includegraphics[width=6.8cm]{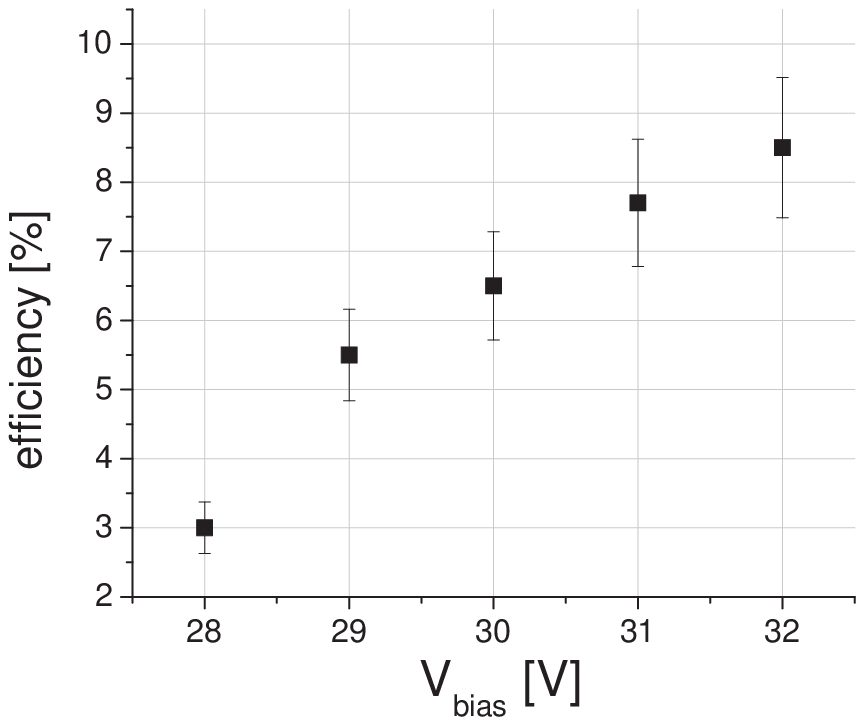}~~
\includegraphics[width=7.3cm]{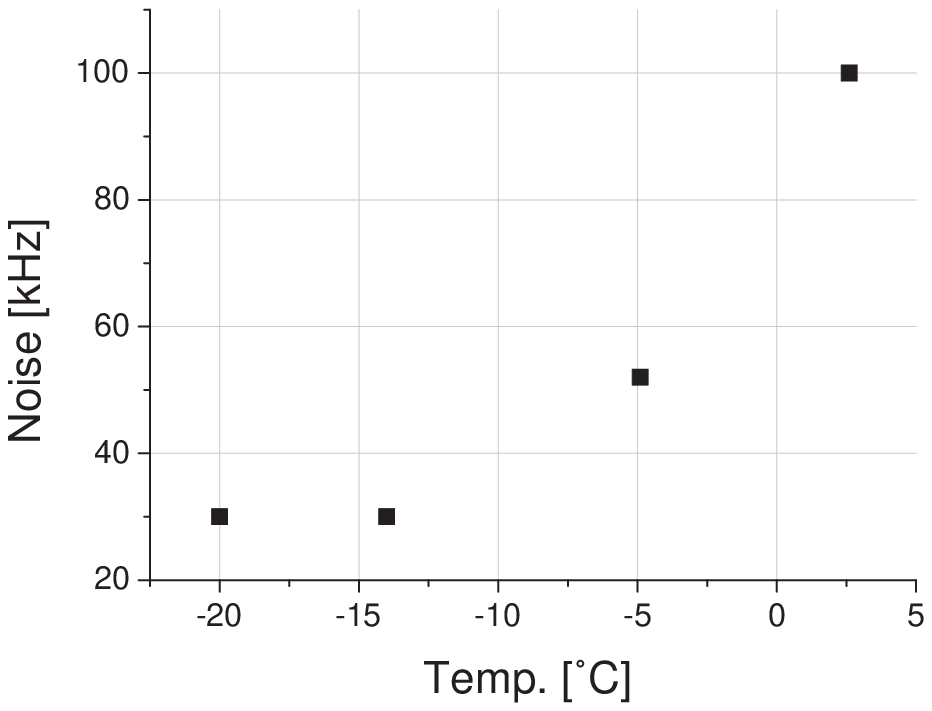}
\caption{Left : Detection efficiency for different values of bias voltage at $T = -7^\circ $ C, when using weak optical pulses of 0.2 photons per pulse at a rate of 430 MHz. Right : Evolution of noise rate for different temperatures with $V_{bias}$ = 32 V. }
\label{VbiasvsEff}
\end{figure}

To measure the noise we operate the device while the light source is switched off. Ambient light contributions are blocked sufficiently to be negligible. Different temperatures, ranging from $+20^\circ$ C to $-20^\circ$ C were applied using the Peltier element.\\
To determine the efficiency we use weak optical pulses. We define efficiency as the probability of getting a SiPM output signal when exacly one photon is incident. We use the highly attenuated light of a mode locked laser (pulse repetition rate = 430 MHz) at $\lambda = 532$ nm with a mean number of photons per pulse of about 0.2 . 
The results of these measurements are shown in Fig.\ref{VbiasvsEff}. The count rates we measure are within 1-3 MHz. As we will see later on, the efficiency is a function of the count rate. The error bars given, are due to the uncertainty introduced by the change of the coupling efficiency when different filters are combined.\\
We find a typical behaviour for the dependencies of a) efficiency vs bias voltage and b) noise vs temperature.  The efficiency increases with bias voltage and reaches a maximum of about 8.5 \% at $V_{bias}= 32$ V. The observed noise decreases with lower temperatures. The minimal noise rate of about 30 kHz, equal to 227 Hz per APD, was found at temperatures below $-14^\circ C$.

\subsection{Counting at 1-40 MHz} 
We start our test of fast counting abilities with a comparison of a state of the art Si-APD (idQuantique) and our SiPM. Considering datasheets, we expect the Si-APD to saturate at about 20 MHz. To prepare pulse frequencies of that order of magnitude, we use a LED at $\lambda = 780$ nm modulated by a periodic block function of 10 ns width and a separation of pulses depending on the applied frequencies. The optical power is chosen sufficiently large ("in saturation") to ensure at least one detection per pulse, that means that a detector with a negligible deadtime -with respect to the applied frequencies- and same efficiency is capable of discriminating between consecutive detections. 
\begin{figure}[t]
\centering
\includegraphics[width=10cm]{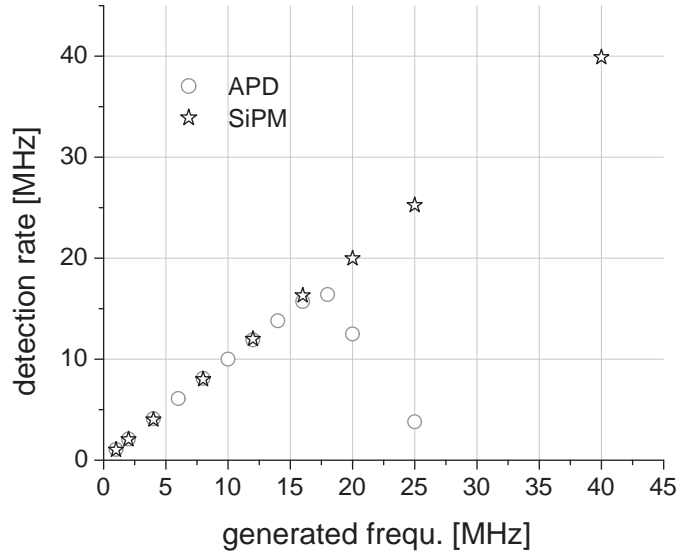}
\caption{Comparison of the count rate for a state of the art Si-APD and our SiPM. While the APD cannot follow the optical pulse rate above 17 MHz, the SiPM reproduces the correct rate up to 40 MHz with an accuracy of about 2\%.}
\label{APDvsSiPM}
\end{figure}
Fig.\ref{APDvsSiPM} shows the responses of both detectors. We find the expected saturation of the Si-APD at about 17 MHz. However, the response of the SiPM continues unaffected reproducing the input pulse rates.  The obtained count rates were accurate within 2\%. This shows a first positive result concerning fast counting with a SiPM.

\subsection{Fast Counting at 430 MHz}
\begin{figure}[h]
\centering
\includegraphics[width=10cm]{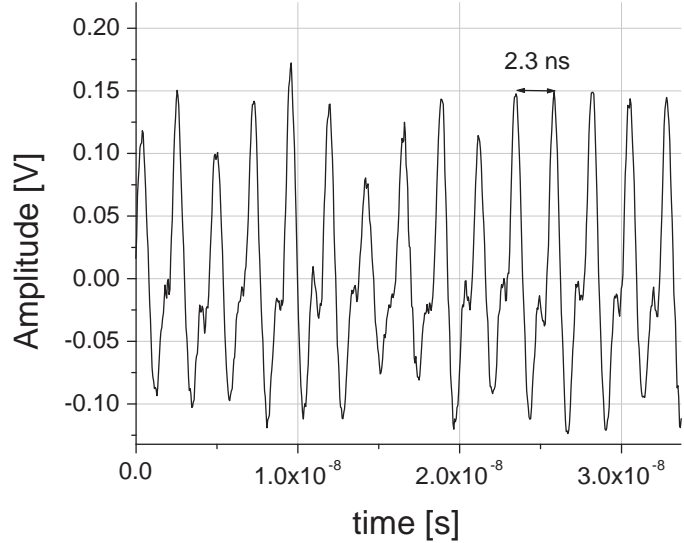}
\caption{Response signal of the SiPM as seen on the oscilloscope. Power per optical pulse is chosen sufficiently large to ensure one detection per pulse. Adjacent count signals are separated sufficiently to be registered correctly.}
\label{FullIllu}
\end{figure}

\begin{figure}[t]
\centering
\includegraphics[width=10cm]{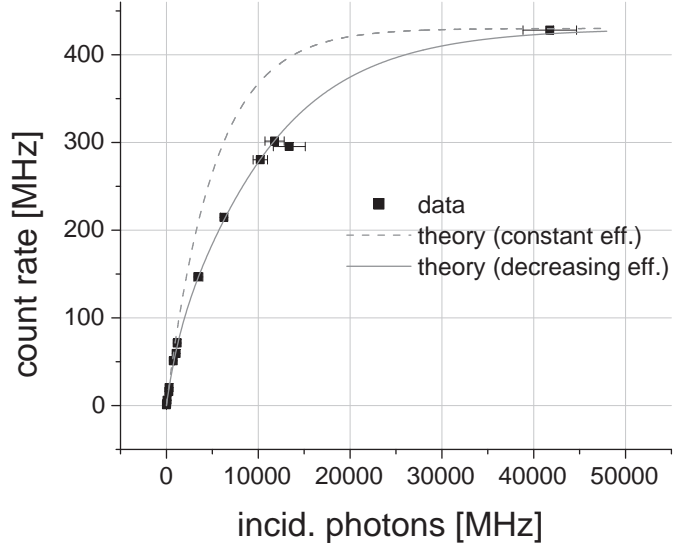}
\caption{Response count rate of the SiPM  to a gradually attenuated mode locked laser with pulse repetition rate of 430 MHz. Rates of incident photons above 430 MHz indicate that a single optical pulse contains more than one photon in average. The data is compared to two theoretical curves. The dashed curve assumes a constant efficiency of $\eta=8.3\%$, fitting data only for very low count rates ($\leq 3$ MHz). The solid curve assumes an exponentially decreasing efficiency.}
\label{LinearPulsed}
\end{figure}

To investigate the behaviour of our SiPM at higher optical pulse rates, we use a mode locked laser (Time-Bandwidth GE-100) at $\lambda = 532$ nm with a pulse repetition rate of 430 MHz and pulse width of about 10 ps. First we test the response in saturation (see  preceding paragraph).  The response of the SiPM as seen on the oscilloscope, is shown in Fig.\ref{FullIllu}. We see well defined single detection signals, separated by a sufficiently large time interval to distinguish between adjacent detections. The count rate found is $f=430$  MHz, thus exacly the frequency of the input signal. This proofs that our setup is  at least capable of detecting consecutive photons separated by only 2.3 ns which corresponds to a frequency of 430 MHz. This resolution is at least 20 times better than the deadtime of common APDs ($\approx$ 50 ns).\\
After that, we gradually attenuate the laser signal to get information about the behaviour of the device on the single photon level.
The result is shown in Fig.\ref{LinearPulsed}. 
The data is compared with a simple detection model taking into account the Poissonian distribution of the number of photons per laser pulse, the statistics of getting a detection when a pulse with a certain number of photons is incident and a certain efficiency $\eta$ to be determined from data. Since we do not discriminate the heights of the detection peaks, but count one detection whenever a certain threshold is crossed, the probability of getting a detection $P_1(\mu)$ is calculated as follows:\\\\
\begin{equation}
P_1(\mu)=\sum_{k\geq 1}^{\infty} p(k,\mu) (1-(1-\eta)^k) =1-p(0,\mu\cdot\eta) = 1-e^{-\mu\cdot\eta} 
\label{equPois}
\end{equation}
where 
$$p(k,\mu):=e^{-\mu}\cdot\frac{\mu^k}{k!} ~~\mbox{(Poissonian distribution)}$$
$$\mu : ~~\mbox{average number of photons per pulse}$$
$$\eta : \mbox{efficiency of detector}$$
The dashed curve (Fig.\ref{LinearPulsed}) assumes a constant efficiency independent of the count rate and incident power, respectively. A value of $\eta=8.3$\% was found by performing a fit for low count rates ($\leq 3$ MHz). As can be seen, this fit overestimates the data for higher rates. 
Taking a closer look at our signal on the oscilloscope, we find a random oscillation of the signal peak heights, ranging from the expected value above 110 mV down to the noise level.  Therefore, it is not obvious where to place the threshold. The higher the optical power the higher the relative number of peaks failing our threshold criteria at 40 mV. Thus we get a lower efficiency and a deviation from the dashed curve.\\
This behaviour is due to limitations of the currently used electronics. In particular we suspect the amplifier to saturate in this region.  
Nevertheless we find a unique correlation between incident optical power and count rate. 
This provides the possibility of using the SiPM as a very sensitive power meter with a dynamic range starting from sub Picowatts to several Nanowatts. For example, a rate of incident photons equal to 5 GHz corresponds to a power of 1.9 nW. To determine the minimal power, we consider a count rate twice as high as the noise level at T$=-7^\circ$ which is 50 kHz. A detection rate of 100 kHz corresponds to a rate of incident photons of about 1200 kHz ($\eta\approx8$\% in this regime). This rate is equal to a power of 0.4 pW.\\
 A good description of the functional dependence can be obtained by assuming an exponential decreasing efficiency
$$\eta(\mu):=p_1 e^{-~p_2\cdot\mu}+p_3$$
The result is shown in Fig.\ref{LinearPulsed} as solid curve, with parameters $(p_1,p_2,p_3)=(0.03,0.157,0.044)$.

\subsection{CW source}

\begin{figure}[h]
\centering
\includegraphics[width=10cm]{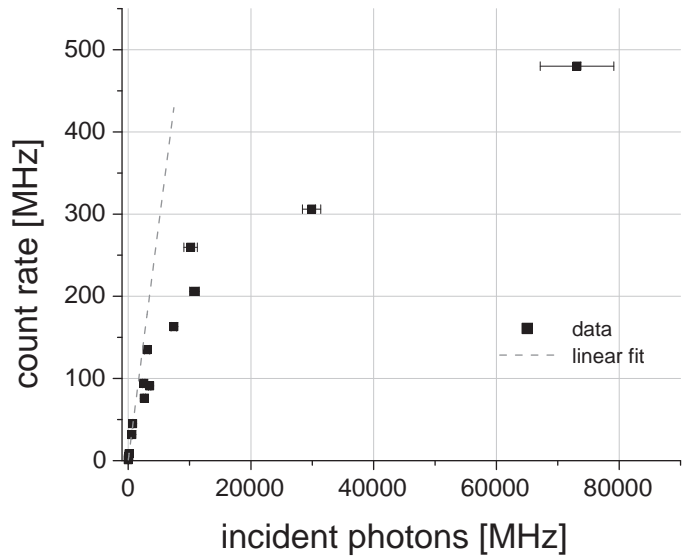}
\caption{SiPM response to a continuous light source. The count rate depends linearly on the number of incident photons up to 50 MHz.}
\label{LinCW}
\end{figure}

As a last test, we illuminate the SiPM with a continuous laser source. We use a HeNe laser (Spectra Physics, $\lambda = 633$ nm) and attenuate it gradually. Here, we expect a lower count rate for a given rate of incident photons with respect to the pulsed case. This is due to the circumstance that the difference in time of arrival of consecutive photons can get smaller than the width of the SiPM output signal (Fig.\ref{singlesignal} (right)). But the maximal count rate should be closer to our expected limit at 500 MHz when the input power is sufficiently high.\\ 
The measurement proceeds exactly in the same way as in the preceding case. The result is shown in Fig.\ref{LinCW}.
We find an almost linear dependence of the count rate from the rate of incident photons up to 50 MHz. The slope of the fit gives a culmulative efficiency of  $\eta = 5.7 $\%.  In the intermediate region we encounter the same problems of electrical amplitude fluctuation, as described before. Applying a large optical power pushes the count rate to 470 MHz.

\section{Conclusion}

We presented an investigation of a SiPM applied for fast counting and multiphoton detection.
In multiphoton detection configuration, sending optical coherent pulses onto the device, we find a good signal discrimination for different numbers of simultaneous detected photons. The obtained results are in good agreement with our model considering additional noise effects due to cross-talk between adjacent APDs.\\ 
The mean part of our work dealt with fast photon-counting abilities of our SiPM. We showed that we are able to considerably outperform deadtime and maximal count rates of photon counting modules based on a single APD. At low optical powers, corresponding to count rates $\leq 3$ MHz, we find an efficiency of $\eta = 8.3 $ \%.  Below  T$ =-14^\circ $ C we measure a moderate dark count rate of 30 kHz which corresponds to an average dark count rate of 227 Hz per APD.\\ We showed that in pulsed mode the device can detect two photons separated by at least 2.3 ns corresponding to 430 MHz. The response of our detector agrees very well with the theory up to a count rate of about 100 MHz. For higher count rates, some electrical limitations (saturation of amplifier) lead to a decreasing efficiency, but it is still possible to predict the count rate for each value of incident photons. 
 
\section*{Acknowledgments}
Financial support from the Swiss Federal Department for Education and Science (OFES) in the framework of the European COST299 project and from the Swiss NCCR "Quantum photonics" is acknowledged.


\end{document}